\documentclass[prd,preprint,tightenlines,superscriptaddress,floatfix,showpacs,preprintnumbers,nofootinbib,eqsecnum]{revtex4}

 \usepackage[dvips,final]{graphicx}
     \usepackage{epsfig}
      \usepackage{bm}

\begin{document}

\thispagestyle{empty} \preprint{\hbox{}} \vspace*{-10mm}

\title{Double -- photon exclusive processes with heavy quark -- heavy antiquark pairs
in high-energy Pb-Pb collisions at LHC}

\author{M.~K{\l}usek-Gawenda}
\email{mariola.klusek@ifj.edu.pl}

\affiliation{Institute of Nuclear Physics PAN, PL-31-342 Cracow, Poland}

\author{A.~Szczurek}
\email{antoni.szczurek@ifj.edu.pl}

\affiliation{Institute of Nuclear Physics PAN, PL-31-342 Cracow, Poland}
\affiliation{University of Rzesz\'ow, PL-35-959 Rzesz\'ow, Poland}

\author{M.V.T. Machado}
\email{magnus@if.ufrgs.br}

\affiliation{High Energy Physics Phenomenology Group, GFPAE IF-UFRGS \\
Caixa Postal 15051, CEP 91501-970, Porto Alegre, RS, Brazil}

\author{V.G. Serbo}
\email{serbo@math.nsc.ru}

\affiliation{Novosibirsk State University, Pirogova 2, 630090,
Novosibirsk, Russia}

\date{November 4, 2010}

\begin{abstract}
The cross section for exclusive heavy quark and heavy antiquark
pair ($Q \bar Q$) production in peripheral ultrarelativistic heavy
ion collisions is calculated for the LHC energy $\sqrt{s_{NN}}$ =
5.5 TeV. Here we consider only processes with photon--photon
interactions and omit diffractive contributions. We present
results in the impact parameter equivalent photon approximation
(EPA) and compare some of them with results obtained by
exact calculations of the Feynman diagrams in the momentum space.
We include both $Q \bar Q$, $Q \bar Q g$ and $Q \bar Q q \bar q$ final states 
as well as photon single-resolved components.
Realistic charge densities in nuclei were taken in the calculation. 
The different components give contributions of the same order of magnitude
to the nuclear cross section.
The cross sections found here are smaller than those for the diffractive 
photon-pomeron mechanism and larger than diffractive pomeron-pomeron 
discussed in the literature.
\end{abstract}


\maketitle

\section{Introduction}

Heavy quark -- heavy antiquark production was studied in the past in
photon-photon, photon-proton, inclusive and exclusive proton--proton
and heavy ion collisions when the ions break apart.
In principle, the heavy quark -- heavy antiquark pairs can
be produced also in exclusive coherent $\gamma \gamma$ processes
when the nuclei stay intact. These processes, in analogy to
lepton pair production \cite{Baur, Serbo}, should be ``increased'' by
the large charges of the colliding nuclei. On the other hand
the effects of the nuclear form factors 
which diminish contributions of large energy
virtual photons should be included.
Recently two of us (M.K.+A.S.) have shown \cite{KS2010} that the
inclusion of realistic form factors (corresponding to realistic charge densities) is
essential for reliable estimation of the cross sections for exclusive
lepton--pair production especially for large rapidities and
large transverse momenta of the produced particles.
The inclusion of the realistic form factor should be even
more important for heavy quark -- heavy antiquark pair production.

In contrast to dileptons the $Q \bar Q$ state cannot be directly observed.
In practice one measures rather heavy mesons or electrons from their semileptonic decays.
Then the final states are already complicated due to hadronization process.
Therefore one has to include also different, yet simple,  
partonic states as $Q \bar Q g$ and $Q \bar Q q \bar q$.

Recently one of us (M.M.) with coworkers made an estimation
of the cross sections for diffractive mechanisms \cite{DMM2010}.
It is of interest to make a realistic estimate of the cross section for photon--photon 
mechanism and compare it with the diffractive contribution.

\section{Heavy quark -- heavy antiquark pair production}

\subsection{Photon--photon subprocesses}

\begin{figure}[!h]    
\centering
\includegraphics[width=.22\textwidth]{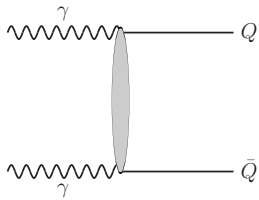}
   \caption{\label{fig:Born}
   \small Representative diagram for the Born amplitude. 
   }
\end{figure}
\begin{figure}[!h]
\includegraphics[width=.6\textwidth]{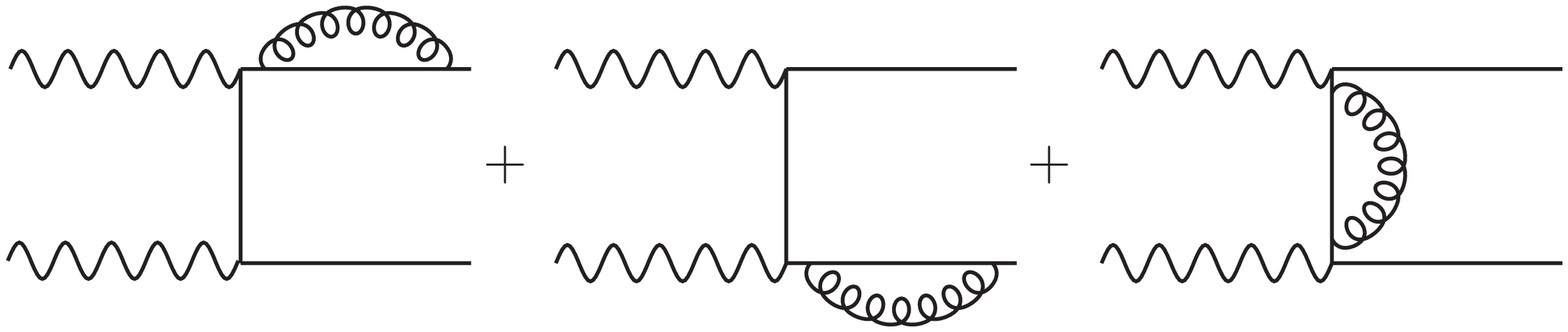}
\includegraphics[width=.6\textwidth]{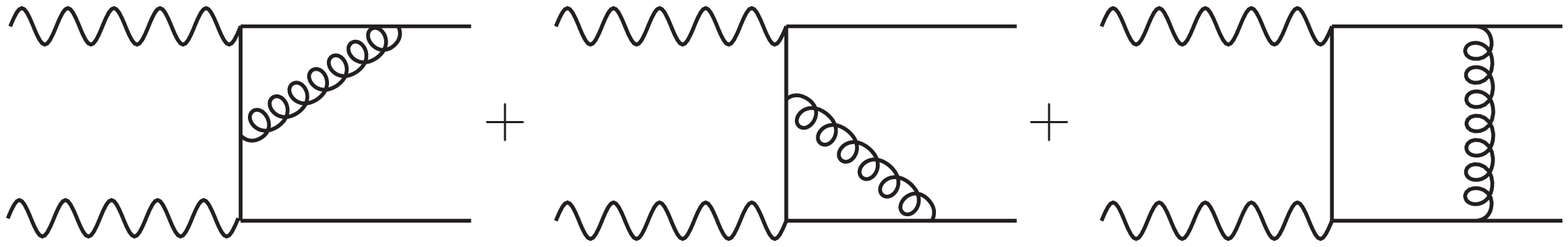}
\includegraphics[width=.6\textwidth]{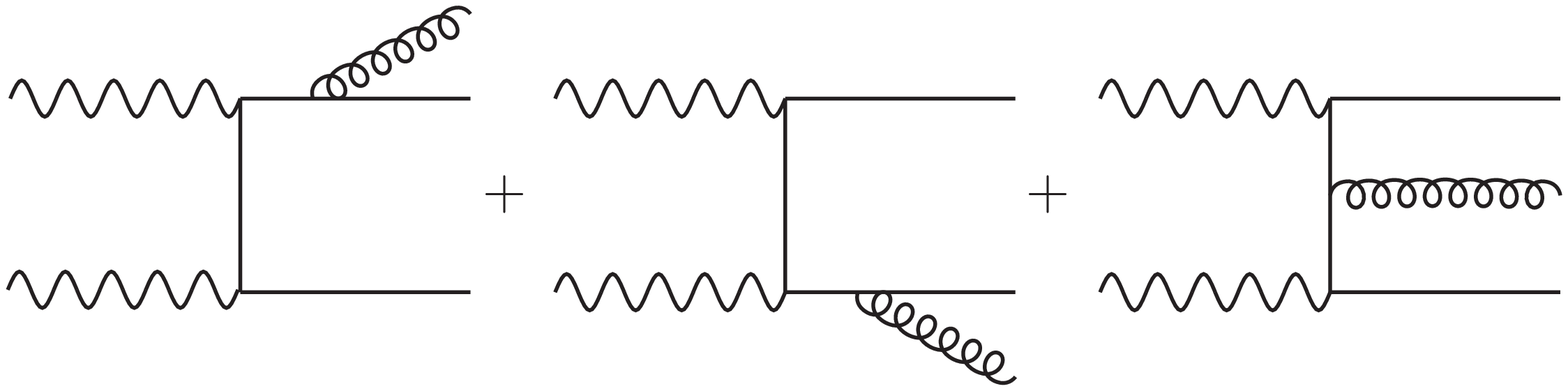}
   \caption{\label{fig:QCD}
   \small Representative diagrams for the leading--order QCD corrections. 
   }
\end{figure}
\begin{figure}[h!]
\begin{minipage}[t]{0.4\textwidth}
\centering
\includegraphics[width=.5\textwidth]{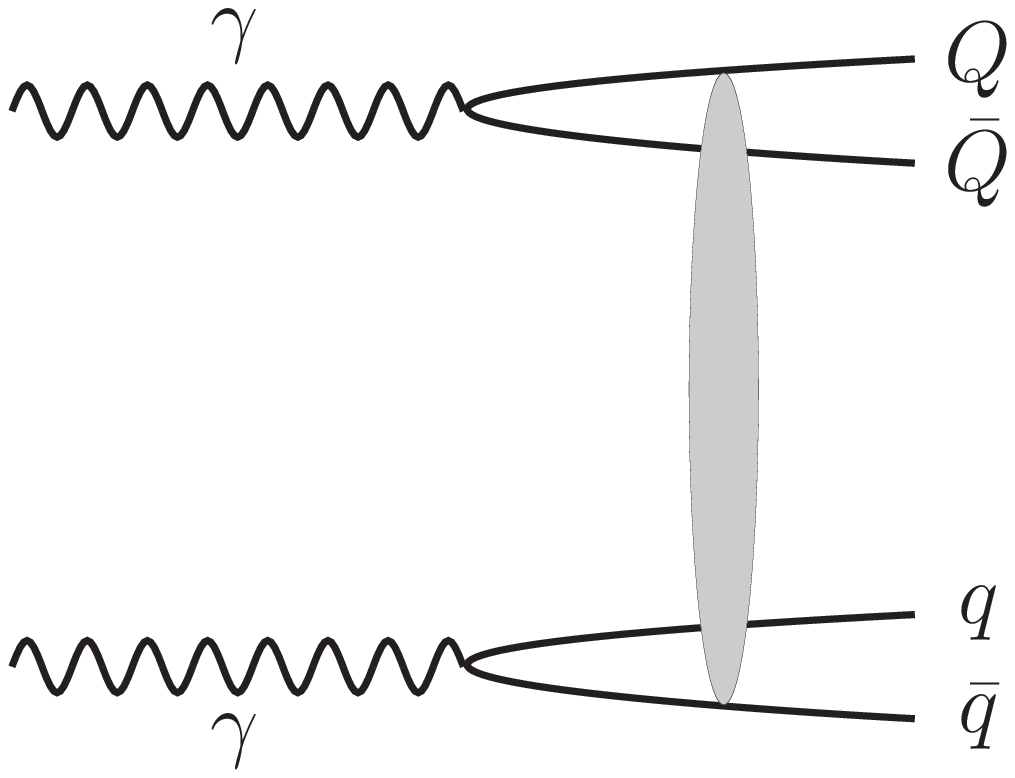}
\end{minipage}
\hspace{0.01\textwidth}
\begin{minipage}[t]{0.4\textwidth}
\centering
\includegraphics[width=.5\textwidth]{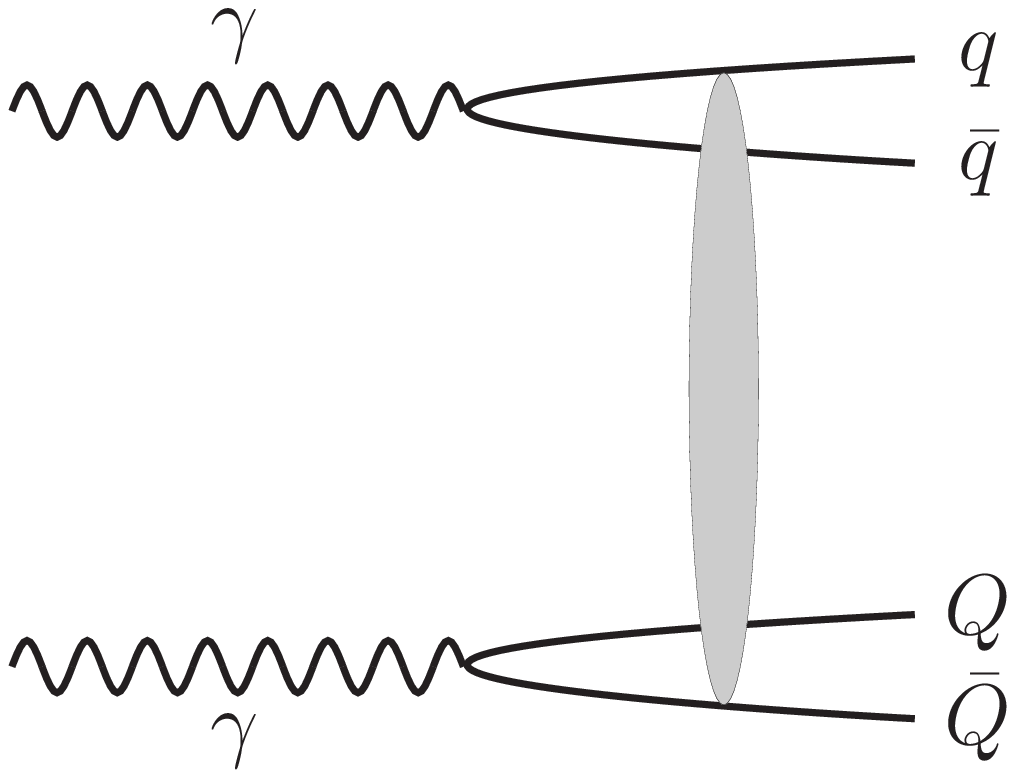}
\end{minipage}
   \caption{\label{fig:QQqq}
   \small Representative diagrams for $Q \bar Q q \bar q$ production. 
   }
\end{figure}
\begin{figure}
\begin{minipage}[t]{0.4\textwidth}
\centering
\includegraphics[width=.5\textwidth]{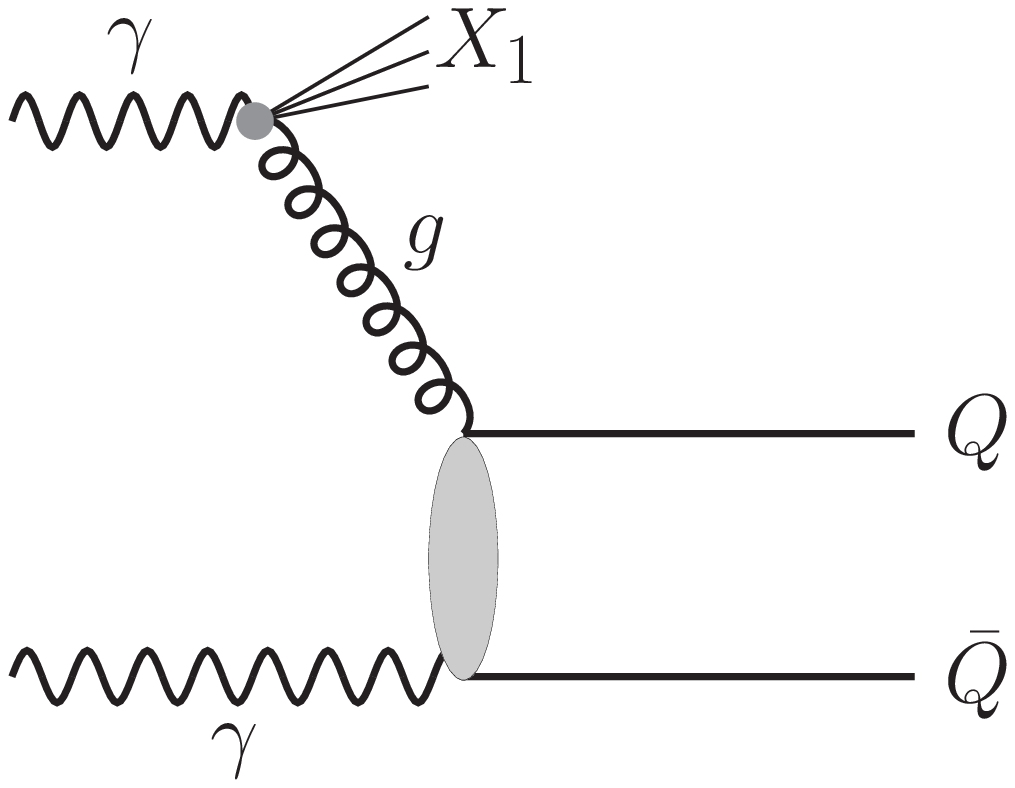}
\end{minipage}
\hspace{0.01\textwidth}
\begin{minipage}[t]{0.4\textwidth}
\centering
\includegraphics[width=.5\textwidth]{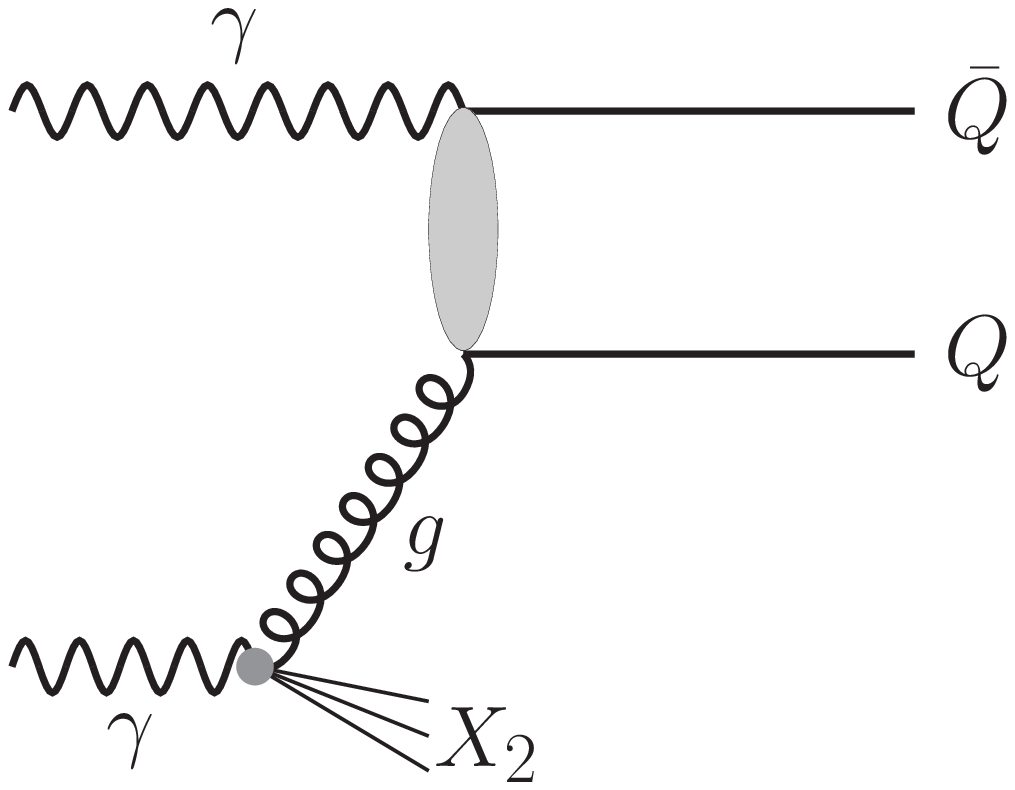}
\end{minipage}
   \caption{\label{fig:sr}
   \small Representative diagrams for the single-resolved mechanism. 
   }
\end{figure}

In the present analysis we include the mechanisms shown in Fig. 
\ref{fig:Born}, \ref{fig:QCD}, \ref{fig:QQqq}, \ref{fig:sr}. 
The first one (called sometimes direct) is
identical as the one for the production of charged lepton pairs.
In contrast to the dilepton production in the case of the quark
production one has to include also QCD corrections.
Corresponding diagrams for NLO approximation are shown in Fig.~\ref{fig:QCD}.
Heavy quarks can be also produced in association with light
quark-antiquark pairs as shown in Fig. \ref{fig:QQqq}.
The last two diagrams correspond to single-resolved components 
when only small part of one photon interacts with the other photon.
All these processes were studied in detail in
Refs.\cite{TKM,szczurek,GM03}.

Let us start with the Born direct contribution. The leading--order
elementary cross section for $\gamma \gamma \to Q \bar{Q}$ at
2-photon energy $W_{\gamma \gamma}$ takes the simple
form
\begin{eqnarray}
  \sigma_{\gamma \gamma \to Q \bar{Q} }^{direct}(W_{\gamma \gamma})
&  = & N_c e_Q^4 \frac{4 \pi \alpha_{em}^2}{W_{\gamma \gamma}^2}  \\
 & \times &
 \left\{ 2  \ln \left[ \frac{W_{\gamma \gamma}}{2m_Q}
\left( 1 +  v \right) \right]
  \left( 1+ \frac{4 m_Q^2 W_{\gamma \gamma}^2 -  8 m_Q^4}
{W_{\gamma \gamma}^4} \right)
- \left( 1 + \frac{4m_Q^2 W_{\gamma \gamma}^2}{W_{\gamma \gamma}^4} \right) v \right\}\, , \nonumber
\label{eq:gg_QQ}
\end{eqnarray}
where $Q \bar{Q} = c \bar{c}, b \bar{b}$, $N_c=3$ is the
number of quark colors, $v = \sqrt{1-\frac{4m_{Q}^2}{W_{\gamma
\gamma}^2}}$ and $e_Q$ is the fractional charge of the heavy
quark. The formula has been derived for the first time in Ref. \cite{BKT1971}.
In the current calculation we take the following heavy quark
masses: $m_c=1.5$ GeV, $m_b=4.75$~GeV.
This formula can be directly used in the impact-parameter-space 
(called here $b$-space for brevity) equivalent
photon approximation (EPA), as we will see below. It is obvious
that the final $Q \bar Q$ state cannot be observed experimentally
due to the quark confinement and rather heavy mesons have to be
observed instead. It is noticed that the presence of additional
few light mesons is rather natural. \footnote{There are also final
states with exclusively two $Q \bar q$ and $\bar Q q$ mesons
\cite{LS2010}. The corresponding cross sections are however much
smaller.} This forces one to include more complicated final states.

In contrast to QED production of lepton pairs in photon-photon
collisions, in the case of $Q \bar Q$
production one needs to include also higher-order QCD processes
which are known to be rather significant.
Here we include leading--order corrections only for the dominant,
in heavy-ion collisions, direct contribution.
The details concerning the higher-order corrections to heavy quark and
heavy antiquark production in photon-photon collisions can be found
in \cite{KMS93,DKZZ93,KL96,KM98,FKL00,Ch04,KMR06,KKMV09}.
In $\alpha_s$-order there are one-gluon bremsstrahlung diagrams
($\gamma \gamma \to Q \bar Q g$) and interferences of the Born
diagram with self-energy diagrams (in $\gamma \gamma \to Q \bar Q$)
and vertex-correction diagrams (in $\gamma \gamma \to Q \bar Q$).
The relevant diagrams are shown in Fig.\ref{fig:QCD}.
In the present analysis we follow the approach presented in Ref. \cite{KKMV09}.
The QCD corrections can be written as
\begin{equation}
  \sigma_{\gamma \gamma \to Q \bar{Q} \left( g \right) }^{QCD}(W_{\gamma \gamma})= 
  N_c e_Q^4 \frac{2 \pi \alpha_{em}^2}{W_{\gamma \gamma}^2} C_F \frac{\alpha_s}{\pi} f^{\left( 1 \right)}.
\label{eq:QCD}
\end{equation}
The function $f^{\left( 1 \right)}$ is calculated using a code
provided by the authors of Ref. \cite{KKMV09}
which uses program package HPL \cite{HPL}. 
In the present analysis the scale of $\alpha_s$ is fixed at $\mu^2=4m_Q^2$.

We include also the subprocess $\gamma \gamma \to Q
\bar{Q} q \bar{q}$, where $q$ ($\bar q$) are light, $u,\, d,\, s$,
quarks (antiquarks). The cross section for this mechanism can be
easily calculated in the color dipole framework
\cite{TKM,szczurek,GM03}. In the dipole--dipole approach
\cite{szczurek} the total cross section for the $\gamma \gamma \to Q
\bar{Q}$ production can be expressed as
\begin{eqnarray}
\sigma^{4q}_{\gamma \gamma \to Q \bar{Q}} \left( W_{\gamma \gamma} \right) & = &
\sum_{f_2 \neq Q} \int \left \vert \Phi ^{Q \bar{Q}} \left( \rho_1, z_1 \right) \right \vert^2 \left \vert \Phi ^{f_2 \bar{f_2}} \left( \rho_2, z_2 \right) \right \vert^2
\sigma_{dd} \left( \rho_1, \rho_2, x_{Qf} \right) d^2 \rho_1 dz_1 d^2 \rho_2 dz_2 \nonumber \\
& + & \sum_{f_1 \neq Q} \int \left \vert \Phi ^{f_1 \bar{f_1}} \left( \rho_1, z_1 \right) \right \vert^2 \left \vert \Phi ^{Q \bar{Q}} \left( \rho_2, z_2 \right) \right \vert^2
\sigma_{dd} \left( \rho_1, \rho_2, x_{fQ} \right) d^2 \rho_1 dz_1 d^2 \rho_2 dz_2, \nonumber \\
\label{sig_dd}
\end{eqnarray}
where $\Phi ^{Q \bar{Q}} \left( \rho,z \right)$ are the quark -- antiquark
wave functions of the photon in the mixed representation and
$\sigma_{dd}$ is the dipole--dipole cross section. Eq.(\ref{sig_dd}) is correct at sufficiently high energy
$W_{\gamma \gamma} \gg 2m_Q$.
At lower energies, the proximity of
the kinematical threshold is a concern.
In Ref. \cite{TKM} a phenomenological saturation--model inspired parametrization
for the azimuthal angle averaged dipole--dipole cross section has been proposed:
\begin{equation}
\sigma^{a,b}_{dd} = \sigma^{a,b}_0 \left[ 1- \exp \left( -
\frac{r^2_{\rm eff}}{4R^2_0 \left( x_{ab} \right)} \right)
\right].
\end{equation}
Here, the saturation radius is defined as
\begin{equation}
R_0 \left( x_{ab} \right) = \frac{1}{Q_0} \left( \frac{x_{ab}}{x_0} \right)^{-\lambda/2}
\end{equation}
and the parameter $x_{ab}$ which controls the energy dependence was given by
\begin{equation}
x_{ab} = \frac{4m_a^2 + 4m_b^2}{W_{\gamma \gamma}^2}.
\end{equation}
In the numerical calculations we are using the model parameters
($x_0,\,\lambda,\,\sigma_0,\,m_q$) for an effective radius $r_{\rm
eff}^2= (\rho_1\rho_2)^2/(\rho_1+\rho_2)$ \cite{TKM} .
Some other parametrizations of the dipole-dipole cross section 
were discussed in the literature (see e.g. \cite{GKCCN2010}).
The cross section for the $\gamma \gamma \to Q \bar Q q \bar q$ 
process here is much bigger than the one corresponding 
to the tree-level Feynman diagram \cite{Ch04} as it effectively 
resums higher-order QCD contributions.

As discussed in Ref. \cite{szczurek} the $Q \bar Q q \bar q$ component 
have very small overlap with the single-resolved component 
because of quite different final state, 
so adding them together does not lead to double counting.
The cross section for the single-resolved contribution can be written as:
\begin{equation}
\sigma_{1-res} \left( s \right) = 	\int d x_1 \left[g_1 \left( x_1, \mu^2 \right) \hat{\sigma}_{g \gamma} \left( \hat{s} = x_1 s \right) \right] +
									\int d x_2 \left[g_2 \left( x_2, \mu^2 \right) \hat{\sigma}_{\gamma g} \left( \hat{s} = x_2 s \right) \right],
\end{equation}
where $g_1$ and $g_2$ are gluon distributions in photon $1$ or photon $2$ 
and $\hat{\sigma}_{q \gamma}$ and $\hat{\sigma}_{\gamma g}$ are elementary cross sections. 
In our evaluation we take the gluon distributions from Ref. \cite{GVR1992}.

\begin{figure}[!h]           
\begin{minipage}[t]{0.325\textwidth}
\centering
\includegraphics[width=1\textwidth]{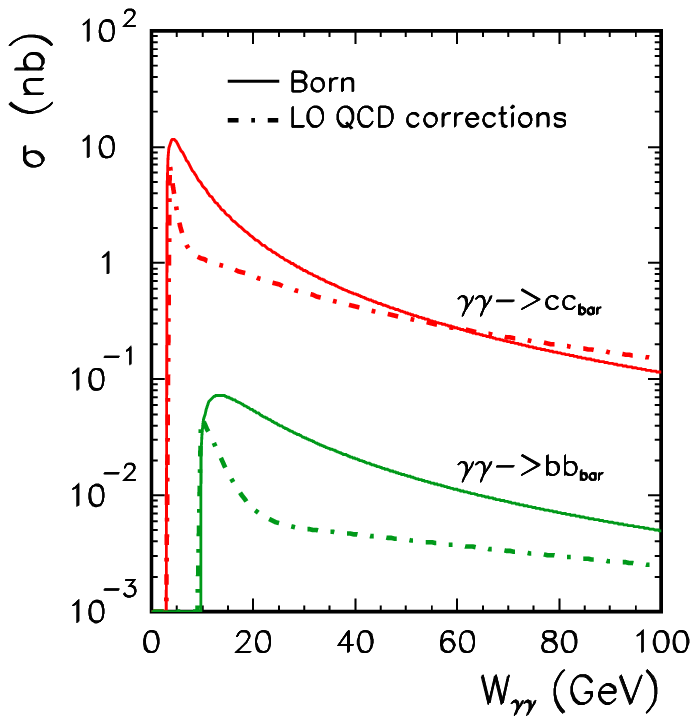}
\end{minipage}
\begin{minipage}[t]{0.325\textwidth}
\centering
\includegraphics[width=1\textwidth]{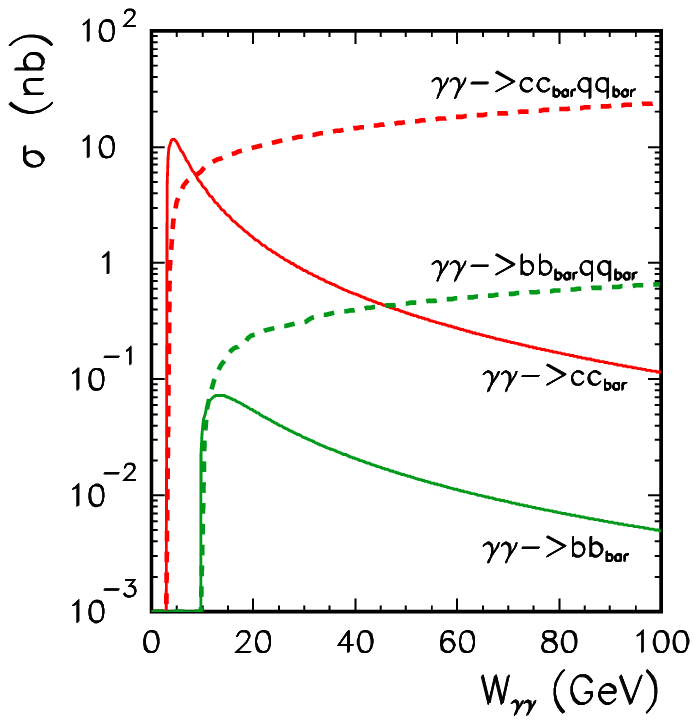}
\end{minipage}
\begin{minipage}[t]{0.325\textwidth}
\centering
\includegraphics[width=1\textwidth]{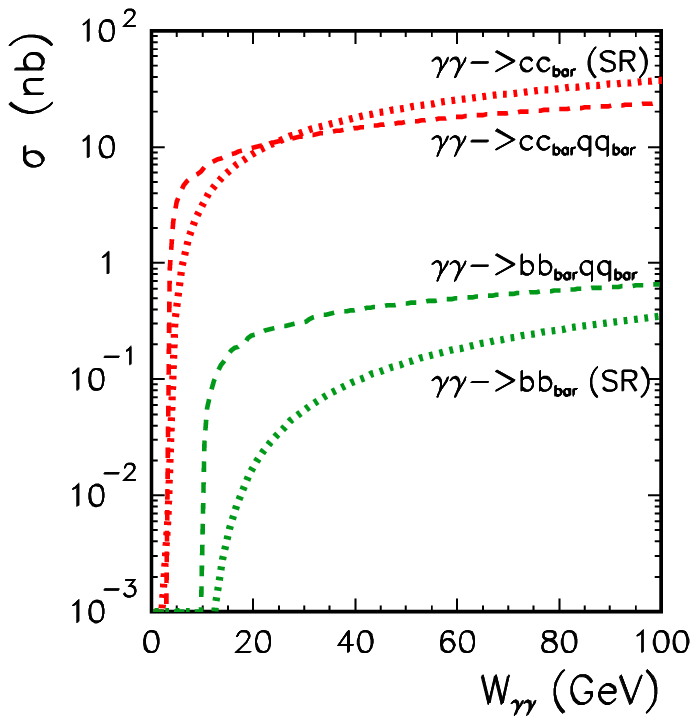}
\end{minipage}

   \caption{\label{fig:el_cross_section}
   \small The elementary cross section for the different processes as a
function of the photon--photon center-of-mass energy. In the left panel we show
the Born cross section (solid line) and leading--order QCD corrections (dash-dotted line).
In the middle panel we show contribution of four-quark final states as calculated
in the saturation model. In the right panel we show in addition contribution of single-resolved
processes.
}
\end{figure}

In Fig. \ref{fig:el_cross_section} we show the elementary cross sections
for all processes as a function of the photon--photon center-of-mass energy.
For charmed quark the direct term dominates at low
energies near the threshold while the four--quark component 
at slightly larger energies and the resolved components at even larger energies.
For bottom quarks the four--quark component is always larger than the direct term.

It is not clear a priori how this will change in the nucleus--nucleus
collisions where one should take into account photon--photon
luminosities. This will be discussed in the result section.

\subsection{$b$-space EPA}

Here we wish to sketch the $b$-space EPA used in the present
analysis. The details on its development can be found in Ref.
\cite{KS2010}.

The total cross section for the $PbPb \rightarrow Pb Pb Q \bar{Q}  $
process can be factorized into the equivalent photon spectra,  $N \left( \omega, b \right)$, and the
$\gamma \gamma \to Q \bar{Q}$ subprocess cross section as:
\begin{eqnarray}
 \sigma\left(Pb Pb \rightarrow Pb Pb Q \bar{Q} ; s_{PbPb}\right)  & = &
\int
{\hat \sigma}\left(\gamma\gamma\rightarrow Q \bar{Q};
W_{\gamma \gamma}  \right) \,
\theta \left(|{\bf b}_1-{\bf b}_2|-2R_A \right)  \nonumber \\
 &\times & N\left(\omega_1,{\bf b}_1 \right)
N\left(\omega_2,{\bf b}_2 \right)
  {\rm d^2}{\bf b}_1
 {\rm d^2}{\bf b}_2 {\rm d}\omega_1 {\rm d}\omega_2 \; .
\label{eq.tot_cross_section}
\end{eqnarray}
After performing a change of integration variables,
the cross section can be expressed as the five--fold integral:
 \begin{eqnarray}
\sigma \left(Pb Pb \rightarrow Pb Pb Q \bar{Q} ; s_{PbPb}\right)
= \int  {\hat \sigma}\left(\gamma\gamma\rightarrow Q \bar{Q}; W_{\gamma \gamma}
\right) \theta \left(|{\bf b}_1-{\bf b}_2|-2R_A \right)& &
\nonumber \\
\times   N \left(\omega_1,{\bf b}_1 \right) N\left(\omega_2,{\bf
b}_2 \right)2 \pi b \, {\rm d} b \, {\rm d} \overline{b}_x \, {\rm
d} \overline{b}_y \frac{W_{\gamma \gamma}}{2} {\rm d}W_{\gamma
\gamma} {\rm d} Y & \, & ,
 \label{eq.tot_cross_section_our}
\end{eqnarray}
where the quantities $\overline{b}_x \equiv (b_{1x}+b_{2x})/2$,
      $\overline{b}_y \equiv (b_{1y}+b_{2y})/2$ and
${\bf b} = {\bf b}_1 - {\bf b}_2$ have been introduced. Eq.
(\ref{eq.tot_cross_section_our}) is used to calculate the total
cross section for the $Pb Pb \to Pb Pb Q \bar{Q}$ reaction as well
as the distributions in impact parameter $b =|{\bf b}|$,
$W_{\gamma \gamma} = M_{Q \bar{Q}}$ and quark pair rapidity $Y(Q
\bar{Q})=\frac{1}{2} \left( y_{Q} + y_{\bar{Q}} \right)$.
A detailed derivation of formula (\ref{eq.tot_cross_section_our})
can be found in \cite{KS2010}. The photon flux can be expressed in
terms of the charge form factors $F(Q^2)$ as:
\begin{equation}
 N \left( \omega,b \right) = \frac{Z^2 \alpha_{em}}{\pi^2}
 \frac{1}{b^2 \omega}
 \left( \int_0^{\infty} u^2 J_1 \left( u \right) \frac{F \left( Q^2
 \right)}{Q^2}\, du
 \right)^2,\; Q^2=\frac{ \left(\frac{b \omega}{\gamma} \right)^2 +u^2}{b^2},
\label{basic_EPA_flux}
\end{equation}
where $J_1$ is the Bessel function of the first kind, $Q^2=-q^2>0$
and $q$ is the four--momentum of the quasireal photon. The form
factor is the Fourier transform of the nucleus charge
distribution, $\rho(r)$:
\begin{equation}
F(Q^2) = \int \frac{4 \pi}{Q} \rho \left( r \right)
\sin{\left(Q\,r \right)}\, r dr = 1 - \frac{Q^2 \langle r^2
\rangle}{3!} + \frac{Q^4 \langle r^4 \rangle}{5!} +\ldots  \; .
\end{equation}

\section{Results}

Now we will discuss the nuclear cross sections obtained within
$b$-space EPA described shortly above.

\begin{figure}[!h]           
\begin{minipage}[t]{0.46\textwidth}
\centering
\includegraphics[width=1\textwidth]{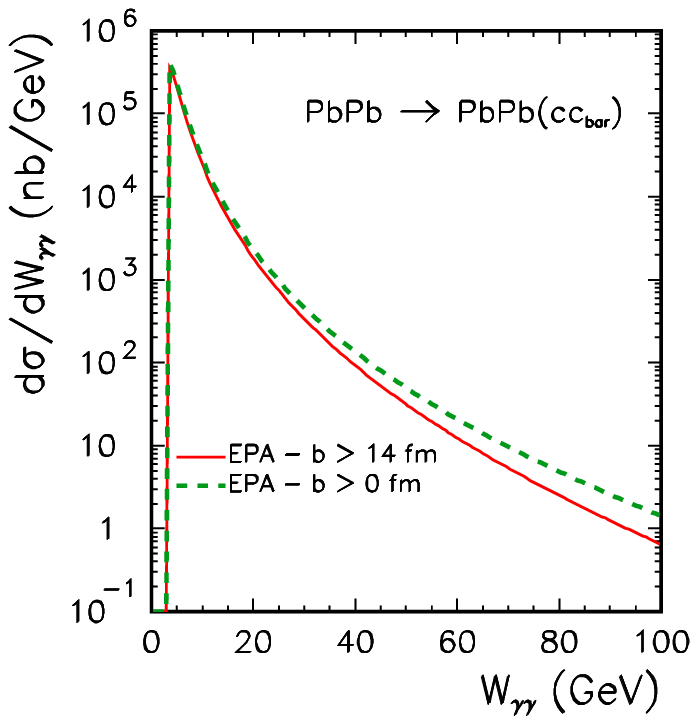}
\end{minipage}
\hspace{0.03\textwidth}
\begin{minipage}[t]{0.46\textwidth}
\centering
\includegraphics[width=1\textwidth]{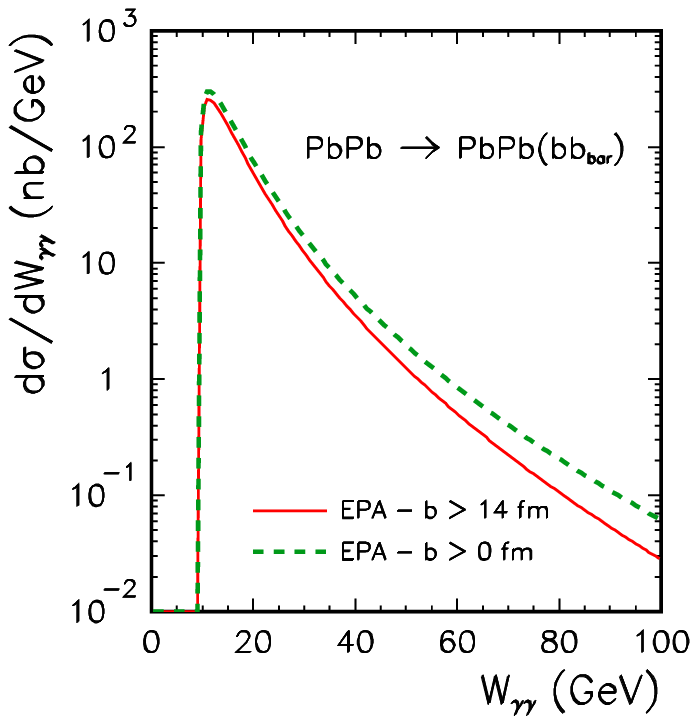}
\end{minipage}
   \caption{\label{fig:dsig_dw_EPA_2b}
The $\gamma\gamma$ subsystem energy distribution, \small $\frac{d
\sigma}{d W_{\gamma \gamma}}$, for $Pb Pb \to Pb Pb c \bar{c}$
(left panel) and $Pb Pb \to Pb Pb b \bar{b}$ (right panel). The
solid line denotes the cross section calculated within EPA approach
for peripheral collisions ($b>14$ fm) while the dashed line
includes also central collisions. }
\end{figure}

Let us start with the direct contribution. In Fig.
\ref{fig:dsig_dw_EPA_2b} we show the differential distribution in
photon--photon energy which for the direct component is also the
distribution in the quark -- antiquark invariant mass. They are
presented without (dashed line) and with (solid line) absorption
effects. This is done in the $b$-space by integrating over impact
parameter either in the full range of $b>0$ (without absorption) or
for $b>R_1 + R_2$ (with absorption), where $R_{1,2}$ are the
nuclei radii. For lead nucleus $R_{A}=1.2\,A^{1/3} \simeq  7$ fm.
The results for the $c \bar c$ are shown on the left hand side
whereas the results for the $b \bar b$ on the right hand side of
the figure. One can clearly see that the cross section for the $c
\bar c$ pair is considerably larger than that for the $b \bar b$
pair (this is because of the charge and mass differences). 
We will return to it when discussing total cross sections at
the end of this section.

\begin{figure}[!h]               
\begin{minipage}[t]{0.46\textwidth}
\centering
\includegraphics[width=1\textwidth]{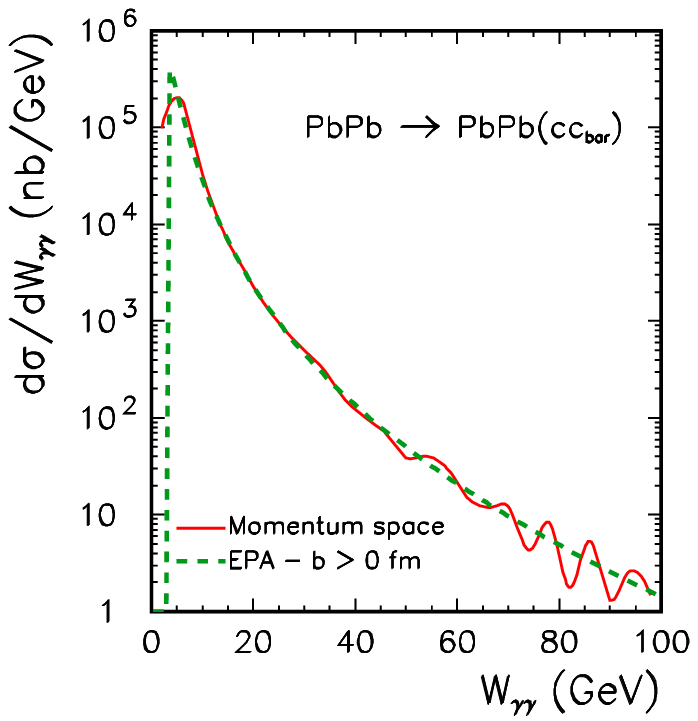}
\end{minipage}
\hspace{0.03\textwidth}
\begin{minipage}[t]{0.46\textwidth}
\centering
\includegraphics[width=1\textwidth]{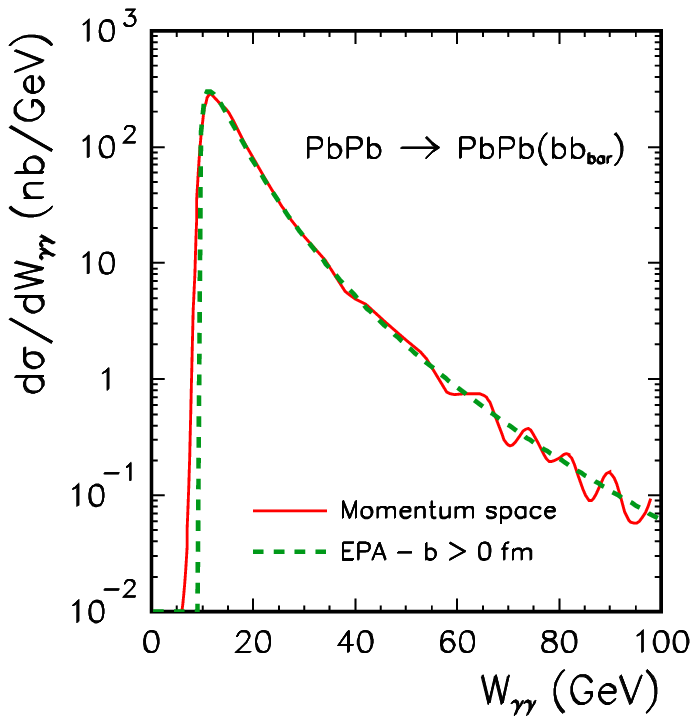}
\end{minipage}
   \caption{\label{fig:dsig_dw_EPA_exact}
   \small The nuclear cross section as a function of 
   $\gamma\gamma$ subsystem energy
for the $Pb Pb \to Pb Pb c \bar{c}$ (left panel) and for the $Pb
Pb \to Pb Pb b \bar{b}$ (right panel) reactions calculated in the
EPA approximation (dashed lines) and in the momentum space (solid
line). }
\end{figure}

Above we have shown results obtained in the equivalent photon
approximation. The same observable can be also obtained
calculating Feynman diagrams in the momentum space. The details of
this calculation have been carefully presented in paper
\cite{KS2010}. In Fig. \ref{fig:dsig_dw_EPA_exact} we compare the
EPA results and those for momentum space calculation. 
The numerical results for the two methods are quite similar.

\begin{figure}[!h]                   
\begin{minipage}[t]{0.46\textwidth}
\centering
\includegraphics[width=1\textwidth]{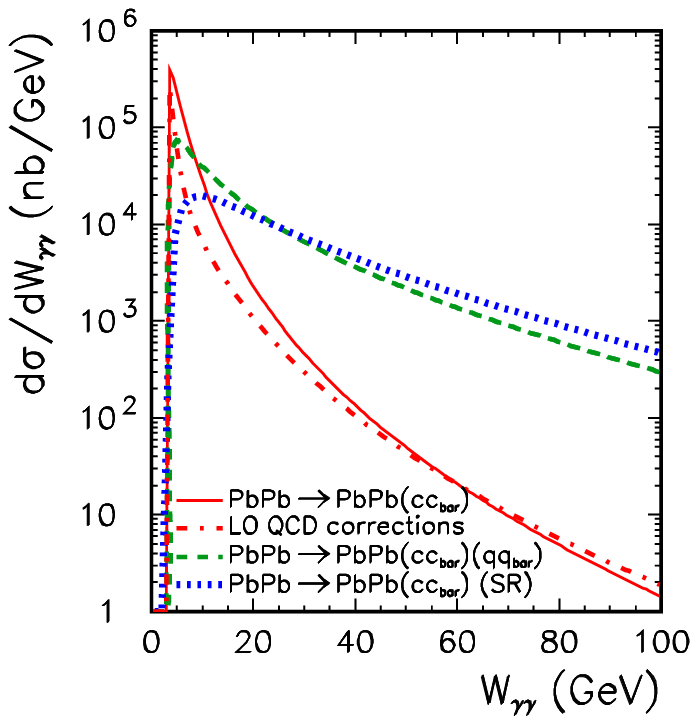}
\end{minipage}
\hspace{0.03\textwidth}
\begin{minipage}[t]{0.46\textwidth}
\centering
\includegraphics[width=1\textwidth]{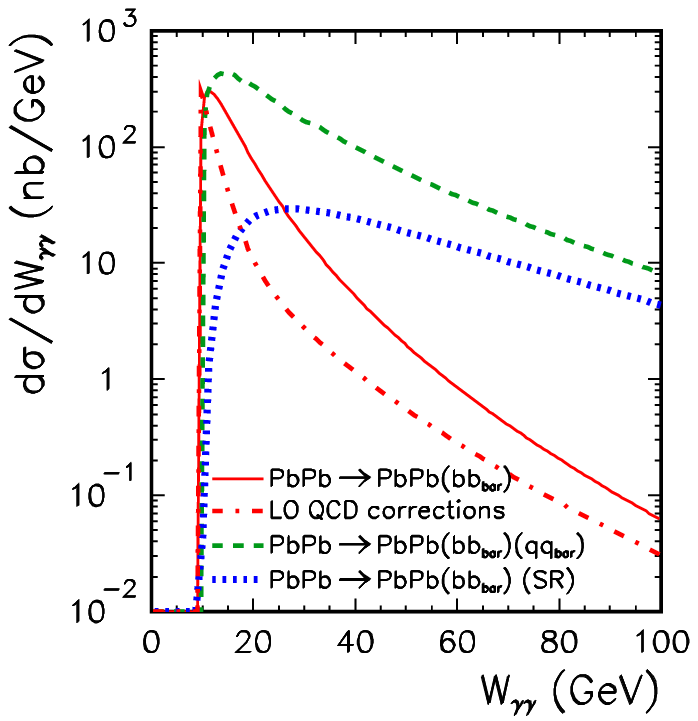}
\end{minipage}
   \caption{\label{fig:dsig_dw_EPA_wc}
   \small The nuclear cross section
as a function of photon--photon subsystem energy $W_{\gamma \gamma}$ 
in EPA. The red solid line denotes the results corresponding to the Born
amplitude ($c\bar{c}$ -left panel and $b\bar{b}$ -right panel). 
The leading--order QCD corrections are shown by the red (on line) dash-dotted line.
For comparison we show the differential
distributions in the case when an additional pair of light quarks
is produced in the final state (dashed lines) 
and for the single-resolved components (dotted line). }
\end{figure}

In Fig. \ref{fig:dsig_dw_EPA_wc} we compare the contributions of
different mechanisms discussed in the present paper
as a function of the photon--photon subsystem energy.
For the Born case it is identical as a distribution
in quark-antiquark invariant mass.
In the other cases the photon--photon subsystem 
energy is clearly different than the $Q \bar Q$ invariant mass. 
Therefore, this distribution is rather
theoretical and would be difficult to measure experimentally.
These distributions reflect the energy dependence of the
elementary cross sections (see Fig.\ref{fig:el_cross_section}).
Please note a sizable contribution of the leading--order corrections
close to the threshold and at large energies for 
the $c \bar c$ case.
Since in this case $W_{\gamma \gamma} > M_{Q \bar Q}$, it becomes
clear that the $Q \bar Q q \bar q$ contributions must have much
steeper dependence on the $Q \bar Q$ invariant mass than the
direct one which means that large $Q \bar Q$ invariant masses are
produced mostly in the direct process. In contrast, small
invariant masses (close to the threshold) are populated dominantly
by the four--quark contribution. Therefore, measuring the
invariant mass distribution one can disentangle the different
mechanisms. As far as this is clear for the $c \bar c$ it is less
transparent and more complicated for the $b \bar b$ production. In
the last case, the experimental decomposition may be in practice
not possible.

\begin{figure}[!h]                   
\begin{minipage}[t]{0.46\textwidth}
\centering
\includegraphics[width=1\textwidth]{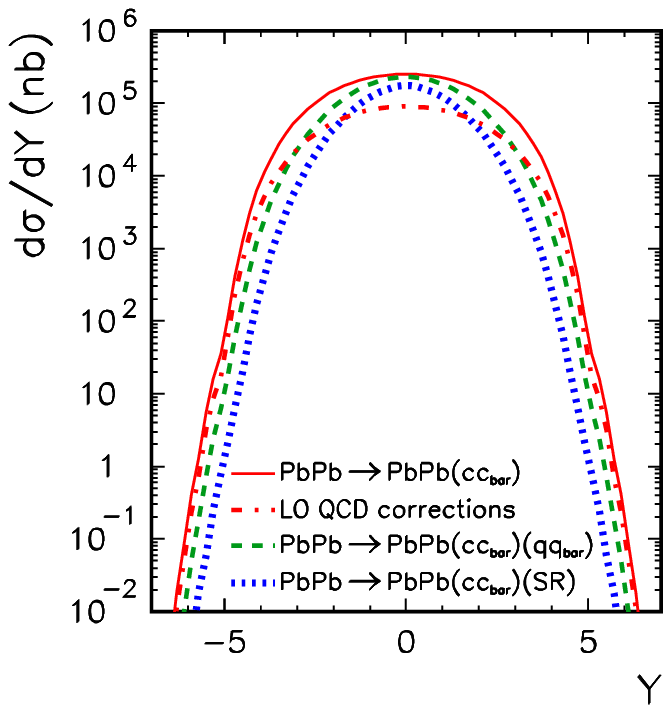}
\end{minipage}
\hspace{0.03\textwidth}
\begin{minipage}[t]{0.46\textwidth}
\centering
\includegraphics[width=1\textwidth]{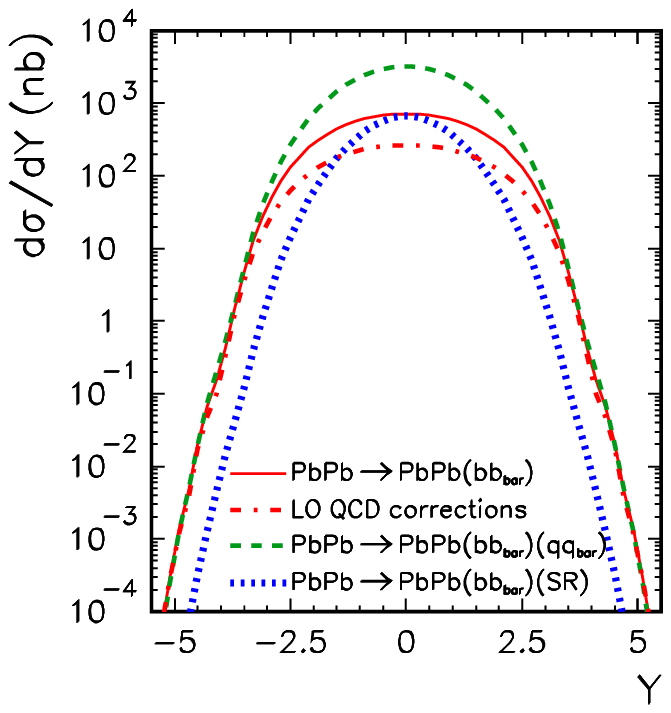}
\end{minipage}
   \caption{\label{fig:dsig_dy_EPA_wc}\textsl{}
   \small The rapidity distribution, $\frac{d \sigma}{d Y}$, in the $b$-space EPA.
   Here the impact parameter is in the whole range of impact parameter ($0<b<\infty$).
   The red (on line) solid line denotes the results corresponding to the Born amplitude
   ($c\bar{c}$ -left panel and $b\bar{b}$ -right panel) and the dash-dotted line corresponds
   to the leading--order QCD corrections.
   For comparison we show the differential distributions in the case
   when an additional pair of light quarks is produced in the final state (dashed lines)
   and for the single-resolved component (dotted line).
}
\end{figure}

Another distribution which can be calculated in the $b$-space EPA
is the distribution in rapidity of the particles produced in the
final state: i.e. rapidity of the $Q \bar Q$ pair for the direct
component, the rapidity of the whole $Q \bar Q g$ or $Q \bar Q q \bar q$ system
for the $Q \bar Q g$ and four--quark component or the rapidity of the very complicated 
system for the single-resolved components. In Fig.~\ref{fig:dsig_dy_EPA_wc} we
present distributions in such a variable for all components,
for $c \bar c$ (left panel) and $b \bar b$ (right panel)
production. It may be quite difficult to reconstruct the 
$Q \bar Q g$ and four--quark rapidity distribution experimentally 
and even more difficult to reconstruct the rapidity 
of the complex final state for the single-resolved components.
The distribution for the single-resolved components is narrower
than for the other components which is related to the fact that 
large part of the energy is taken by the photon remnants. 

Finally, in Tables 1, 2, 3 and 4 we have summarized the results for the
total cross sections for different components calculated within
distinct methods: $b$-space EPA, exact momentum space and
momentum-space EPA described in detail in~\cite{Serbo}.

In Table 1 we present the cross sections for the Born direct component
only. In order to illustrate the absorption effect we show both
the integral calculated from 0 to ``infinity'' and the integral
calculated from $R_1 + R_2$ to ``infinity''. The integration in
$b$ is only slowly convergent, especially for the lighter $c \bar
c$ pairs. Therefore we also show the practical upper limit
dependence of the cross section. We are not sure that the upper
limit for $c \bar c$ is sufficient. The results obtained within
momentum-space EPA \cite{Serbo} are very similar to those obtained
in the exact momentum space method \cite{KS2010}. For $b \bar b$
these results are also in good agreement with the $b$-space EPA.

In Table 2 we have ensambled leading--order QCD corrections. They
constitute about one third of the Born contribution.

In Table 3 we show results for the $Q \bar Q q
\bar q$ components obtained using $b$-space EPA. The cross
sections here are of the similar order of magnitude as those for
the direct component. For charm, the cross section for
$Q\bar{Q}q\bar{q}$ is approximately equal to the direct
contribution. On the other hand, for bottom the result is almost
four times bigger than the direct component. This is due the
dominance of four--quark component even near threshold as shown in
Fig. 5. That feature is already known from Ref. \cite{szczurek},
where the dipole-model was compared with LEP data for $\sigma
(e^+e^- \rightarrow b\bar{b}X)$. 

For completeness in Table 4 
we present results for the single-resolved component.
The cross sections for this component is comparable to the other components.  

Finally, in Table 5
we present cross sections which include all discussed mechanisms
and their relative contributions. We see that the Born mechanism
dominates for $c \bar c$ production 
but four-quark component for $b \bar b $ production.

The event rates should be large
also after hadronization process. For example, the total
$c\bar{c}$ and $b\bar{b}$ two-photon rates in peripheral PbPb
collisions over a $10^6$ s run at LHC are $N(c\bar{c})=10.4 \times 10^5$ 
and $N(b\bar{b})=4.6 \times 10^3$ including
Born, QCD-corrections, single-resolved and four--quark component. 
This is done
using the $b$-space EPA with absorption and taking a luminosity of
${\cal L}_{PbPb}=4.2\times 10^{26}$ cm$^{-2}$s$^{-1}$. The results
in Tables 1 and 4 can be compared directly to previous studies
which rely on QCD collinear factorization approach, for instance
in Ref. \cite{Klein:2002wm}. Notice that for charm the direct
contribution produces similar cross sections compared to
\cite{Klein:2002wm} whereas for bottom it is a factor two bigger.
The $Q\bar{Q}q\bar{q}$ contribution was not included in
\cite{Klein:2002wm} and in addition the single and double resolved
processes were shown to be negligible for LHC energy compared to
the direct one. As a short final comment, the cross sections
presented in Tables 1 and 4 are larger than the estimations for
the exclusive heavy-quark production in double-pomeron exchange
(DPE) process (it was found in Ref. \cite{DMM2010}
$\sigma^{\mathrm{DPE}}_{\mathrm{PbPb}}(c\bar{c})\approx 4.2$
$\mu$b and $\sigma^{\mathrm{DPE}}_{\mathrm{PbPb}}(b\bar{b})\approx
0.2$ $\mu$b) whereas they are smaller than the quark-pair production
in photon-pomeron processes.

\begin{table}

\caption{
Total cross section for the $Q \bar Q$ component calculated within
different methods.}

\begin{tabular}{|c||c|c|c|c|} \hline
 Process                            & \multicolumn{2}{|c|}{$b$-space EPA}                                 & Momentum-         & Momentum-         \\
                                    & b$>$0                     & b$>$14 $fm$                             & -space            & -space EPA        \\ \hline \hline
 $PbPb\to PbPb\,c \overline{c}$     & 1.18 $m b$ (b$<$4000 $fm$)    & 1.05 $mb$ (b$<$4000 $fm$)        & 1.36 $mb$       & 1.230 $mb$       \\
                                    & 1.13 $m b$ (b$<$1000 $fm$)    & 1.00 $mb$ (b$<$1000 $fm$)        &                   &                   \\
 $PbPb\to PbPb\,b \overline{b}$     & 2.53 $\mu b$                 & 2.05 $\mu b$ (b$<$1000 $fm$)      & 2.54 $ \mu b$    & 2.54 $\mu b$         \\ \hline
\end{tabular}

\end{table}
\begin{table}

\caption{The leading--order QCD corrections to the total cross section
within $b$-space EPA.}

\begin{tabular}{|c||c|c|c|} \hline

  Process                               & \multicolumn{2}{|c|}{$b$-space EPA} \\
                                        & b$>$0             & b$>$14 $fm$      \\ \hline \hline

 $PbPb\to PbPb \, c \overline{c}$ & 0.41 $mb$      & 0.36 $mb$   \\ \hline
 $PbPb\to PbPb \, b \overline{b}$ & 1.00 $\mu b$    & 0.83 $\mu b$      \\ \hline

\end{tabular}

\end{table}
\begin{table}

\caption{Total cross section for the $Q \bar Q q \bar q$
components within $b$-space EPA.}

\begin{tabular}{|c||c|c|c|} \hline

  Process                               & \multicolumn{2}{|c|}{$b$-space EPA} \\
                                        & b$>$0             & b$>$14 $fm$      \\ \hline \hline

 $PbPb\to PbPb \, c \overline{c}q \overline{q}$ & 0.82 $mb$      & 0.67 $mb$   \\ \hline
 $PbPb\to PbPb \, b \overline{b}q \overline{q}$ & 9.40 $\mu b$    & 6.98 $\mu b$      \\ \hline

\end{tabular}

\end{table}
\begin{table}

\caption{Total cross section for the single-resolved components within $b$-space EPA.}

\begin{tabular}{|c||c|c|c|} \hline

  Process                               & \multicolumn{2}{|c|}{$b$-space EPA} \\
                                        & b$>$0             & b$>$14 $fm$      \\ \hline \hline

 $PbPb\to PbPb \, c \overline{c}$ 		& 0.52 $mb$      	& 0.39 $mb$   \\ \hline
 $PbPb\to PbPb \, b \overline{b}$ 		& 1.51 $\mu b$    	& 0.97 $\mu b$      \\ \hline

\end{tabular}

\end{table}
\begin{table}

\caption{Partial contributions of different mechanisms.}

\begin{tabular}{|c||c||c|c|c|c|} \hline

                   		& $\sigma_{tot}$	& Born		& QCD-corrections	& 4-quark	& Single-resolved      \\ \hline \hline

 $c \overline{c}$ 		& 2.47   $m b $     & 42.5 \% 	& 14.6 \%			& 27.1 \%	& 15.8 \%				\\ \hline
 $b \overline{b}$ 		& 10.83 $\mu b$    	& 18.9 \%   & 7.7 \%			& 64.5 \%	& 8.9 \%				\\ \hline

\end{tabular}

\end{table}

\section{Conclusions}

In the present paper we have concentrated on production of
heavy quark -- heavy antiquark pairs in coherent photon--photon
subprocesses. A discussion on similar diffractive processes have been already
presented in the literature and will be not repeated here.
The photon--photon processes are dominant in the case of exclusive
production of charged lepton pairs.
In our calculations we have used realistic nuclear form factors
calculated as Fourier transform of the realistic charge density
of the nucleus known from the electron scattering off nuclei.
Recently, this was shown to be crucial for reliable estimation
of the exclusive lepton pair production.

We have calculated cross sections for exclusive production
of charm--anticharm and bottom--antibottom pairs, for the $Q \bar Q g$ and
$Q \bar Q q \bar q$ final states as well as for 
the single-resolved components in the high--energy peripheral
lead--lead collisions for the LHC energy $\sqrt{s}_{NN}$ = 5.5 GeV.
Large cross sections have been found in the case of charm quarks
(antiquarks) production. In contrast to the exclusive dilepton production
in the case of the heavy quark - heavy antiquark production large QCD 
corrections appear. Their fractional contribution strongly depends on
the photon-photon subsystem energy.

Two methods have been used to calculate the Born contribution: 
impact parameter equivalent photon
approximation ($b$-space EPA) and the Feynman diagrammatic
approach for the Born component called here 
momentum space approach. The $b$-space EPA is an approximation 
but allows to include absorption effects in a
simple way by limiting the range of integration over impact
parameter. The direct contribution was calculated in both the
$b$-space EPA and the Feynman graph approach while 
the leading--order QCD correction,
four--quark component as well as single-resolved components 
only in the $b$-space EPA.

We have presented total (phase space integrated) cross sections as
well as some selected differential distributions relatively easy
to calculate in the $b$-space EPA. The absorption effects turned
out to be larger for bottom quarks (20 \%) than for charm quarks
(10 \%). Since both methods lead to similar effects one can use
the momentum space approach to calculate, or at least to estimate,
different observables which are not straightforward in the
$b$-space approach.

We have found that the contributions of two-- and four-- quark
and single-resolved
final states are of similar size. We have found also that the large
invariant masses of the $Q \bar Q$ system are populated dominantly
by the direct $\gamma \gamma \to Q \bar Q$ subprocesses
while smaller invariant masses by
the $\gamma \gamma \to Q \bar Q g$, 
$\gamma \gamma \to Q \bar Q q \bar q$ or single-resolved components.
This could be potentially helpful in experimental identification
of the all components.
There are known experimental methods how to distinguish large transverse
momentum $b$ ($\bar b$) jets, therefore exclusive measurement
of such jets should be possible in the LHC experiments.

The cross sections found for the QED processes discussed in the present
paper seem smaller than those found in diffractive photon-pomeron processes 
but smaller than diffractive pomeron-pomeron process.
The diffractive processes are more difficult to be reliably
calculated. A combined simultaneous analysis of all processes
including different differential distributions seems indispensable
in the future. Since the QED processes, as demonstrated here, can
be reliably calculated it can be used as a background to
the much more involved diffractive processes.

\vspace{1cm}

{\bf Acknowledgments}

We are indebted to Zakaria Merebashvili and Oleg Veretin
for providing a code which calculates QCD corrections
and for a discussion of the related physics.  
This work was partially supported by the Polish grant N N202
078735 and N N202 249235. MTVM was supported by science funding agency CNPq, Brazil.
V.G.S. is supported by of the Russian (RFBR 09-02-00263;
NSh-3810.2010.2) and USA (NSF-PHY-8555454; the Missouri Research
Board) grants.


%

\end{document}